\input harvmac
\newcount\figno
\figno=0
\def\fig#1#2#3{
\par\begingroup\parindent=0pt\leftskip=1cm\rightskip=1cm\parindent=0pt
\baselineskip=11pt
\global\advance\figno by 1
\midinsert
\epsfxsize=#3
\centerline{\epsfbox{#2}}
\vskip 12pt
{\bf Fig. \the\figno:} #1\par
\endinsert\endgroup\par
}
\def\figlabel#1{\xdef#1{\the\figno}}
\def\encadremath#1{\vbox{\hrule\hbox{\vrule\kern8pt\vbox{\kern8pt
\hbox{$\displaystyle #1$}\kern8pt}
\kern8pt\vrule}\hrule}}

\overfullrule=0pt

%
\def\tilde{\widetilde}

\def\Z{{\bf Z}}

\def\S{{\bf S}}
\def\R{{\bf R}}

\font\zfont = cmss10 
\font\litfont = cmr6

\def\bigone{\hbox{1\kern -.23em {\rm l}}}
\def\ZZ{\hbox{\zfont Z\kern-.4emZ}}
\def\half{{\litfont {1 \over 2}}}

\Title{hep-th/9609122, IASSNS-HEP-96/96}
{\vbox{\centerline{ON FLUX QUANTIZATION IN $M$-THEORY}
\bigskip
\centerline{AND THE EFFECTIVE ACTION}}}
\smallskip
\centerline{Edward Witten\foot{Research supported in part
by NSF  Grant  PHY-9513835.}}
\smallskip
\centerline{\it School of Natural Sciences, Institute for Advanced Study}
\centerline{\it Olden Lane, Princeton, NJ 08540, USA}\bigskip

\medskip

\noindent

The quantization law for  the antisymmetric tensor field of $M$-theory
contains a gravitational contribution not known previously.
When it is included, the low energy effective action of $M$-theory,
including one-loop and Chern-Simons contributions, is well-defined.
The relation of $M$-theory to the $E_8\times E_8$ heterotic string
greatly facilitates the analysis.
 
\Date{September, 1996}

\newsec{Introduction}

This paper is devoted to explaining some 
topological points concerning eleven-dimensional
$M$-theory.  Roughly speaking, we will explain one new physical
effect and show how it enters in re-interpreting, or avoiding,
several potential anomalies.  

The new physical effect involves the three-form $C$ of eleven-dimensional
supergravity and its field strength $G=dC$.  It has been believed
that $G$ is constrained precisely by a flux quantization law,
which  says that if $G$ is correctly normalized its periods
are integer multiples of $2\pi$.  
As we will see, this is not quite correct.

For the precise statement, recall first that in $M$-theory
we can assume that the space-time manifold $Y$ is a spin manifold
since spinors exist in the theory.\foot{In this paper, we generally
assume
that $Y$ is orientable, although, as $M$-theory conserves
parity (this is one of the points we will re-examine), one could relax
this.  If $Y$ is unorientable, it should carry
a ``pin'' structure rather than a spin structure.  We do consider
an unorientable example at the end of section two.} 
 For $Y$ a spin manifold, the
first Pontryagin class $p_1(Y)$ is divisible by two in a canonical
way.  We set
\eqn\wella{\lambda(Y)={p_1(Y)\over 2}.}
$\lambda$ in turn may or may not be divisible by two, depending on
the topology of $Y$.  

It turns out that the object that must have integral periods is not
$G/2\pi$  but $G/2\pi-\lambda/2$.  We denote the cohomology class of $G/2\pi$ 
as $[G/2\pi]$ and we describe the situation by saying that
\eqn\unbu{\left[{G\over 2\pi}\right] -{\lambda\over 2} \in H^4(Y,\Z).}
(A perhaps more precise way to say this is that it is not
$[G/2\pi]$ but $[G/\pi]$ that
is well defined as an integral cohomology class, and that this class is
congruent to $\lambda$ modulo 2.)
This shift from  the naive quantization law is obtained  in section two
by a very brief argument using the relation of $M$-theory
to the $E_8\times E_8$ heterotic string \nref\hw{P. Horava and
E. Witten, ``Heterotic And Type I String Dynamics From Eleven 
Dimensions,'' Nucl. Phys. {\bf B460} (1996) 506, hepth/9510209;
``Eleven-Dimensional Supergravity On A Manifold With Boundary,'' 
hepth/9603142.}\hw.  We also give a second (and perhaps
more precise) derivation of the
shift using membrane world-volume anomalies.  The rest of the
paper is devoted to applications of this effect, beginning at the
end of section two with an application to $M$-theory on $(\S^1)^5/\Z_2$.

In section three, we reconsider an effect noted  recently
in compactification of $M$-theory to three dimensions 
\nref\svw{S. Sethi, C. Vafa, and E. Witten,
``Constraints On Low-Dimensional String Compactifications,''
hepth/9606122.}\svw.  It was shown that standard
$M$-theory compactification on $Y=\R^3\times X$, with $X$ an 
eight-manifold, is consistent only if $X$ obeys a certain
topological condition.  We will see that this condition follows
from \unbu.  \unbu\ implies that $G$ can be zero only if
$\lambda(X)$ is divisible by two (in the integral cohomology of $X$).
In \svw\ (as in most discussions of compactification), 
$G$ was set to zero.  An inconsistency therefore arises unless
$\lambda(X)/2$ is integral.  As we will show in section three,
when  $\lambda(X)/2$ is integral, the anomaly found in \svw\ vanishes.
  
What to do when $\lambda(X)/2$ is not integral is now clear.
One must endow $X$ with a suitable non-zero  $G$ with half-integral
periods of $G/2\pi$.  
Compactifications with non-zero $G$ have not been much studied,
but in the case of compactification on a Calabi-Yau four-fold,
there is a very elegant framework \nref\beckers{K. Becker and
M. Becker, ``$M$-Theory On Eight-Manifolds,'' hepth/9605053.}
\beckers\ for
incorporating $G$ in a fashion that preserves supersymmetry.
(See also \ref\duffo{M. J. Duff, B. E. W. Nilsson, and C. N. Pope,
``Spontaneous Supersymmetry Breaking By The Squashed Seven
Sphere,'' Phys. Rev. Lett. {\bf 50} (1983) 2043.} for an older
example of a different sort.) 
Of course, one can turn on (integral) $G$ even if $\lambda/2$
is integral, but when $\lambda/2$ is non-integral, one {\it must}
take $G$ non-zero.

Integrality of $\lambda/2$ has an attractive interpretation
that enters the argument of section three: it is equivalent
to evenness of the intersection form on $H^4(X,\Z)$.

In section four, we move on to a question of which the discussion
in section three is really a special case: defining the
Chern-Simons interactions of the low energy limit of $M$-theory.
Thus, the quantization of $G$ means that the $C\wedge G\wedge G$
interaction of eleven-dimensional supergravity is a kind of Chern-Simons
coupling.  One may wonder whether, given the value of the quantum
of $G$, this coupling is correctly normalized so as to be
single-valued modulo $2\pi$.  The answer is ``no''; it is too
small by a factor of six.    To make
sense of the Chern-Simons interaction requires two ingredients:
gravitational corrections to the $CGG$ coupling; and some congruences.

The gravitational corrections come from two sources:
a term first computed in \nref\vw{C. Vafa and E. Witten,
``A One-Loop Test Of String Duality,''  Nucl. Phys. {\bf B447}
(1995), hep-th/9505053.}\vw\ and
further studied in \nref\duff{M. J. Duff, J. T. Liu, and R. Minasian,
``Eleven-Dimensional Origin Of String/String Duality: A One Loop
Test,'' hepth/9506126.}\duff; and the $\lambda/2$ term
in \unbu.   The necessary congruences are potentially baffling
-- unless one knows the relation of $M$-theory to the $E_8\times E_8$
heterotic string, which gives them  naturally.  
Putting everything together,  we show that the Chern-Simons
couplings are completely well-defined modulo a term that is
independent of $G$.

Section five is devoted to the $G$-independent term, which turns
out to be essential in resolving a longstanding though perhaps
relatively little-known puzzle.
In $8k+3$ dimensions, Dirac-like operators have eigenvalues
which are real but not necessarily positive.  There is sometimes
\ref\aw{L. Alvarez-Gaum\'e and E. Witten, ``Gravitational
Anomalies,'' Nucl. Phys. {\bf B234} (1983) 269;  see p. 309.} 
(see \nref\redlich{A. N. Redlich, ``Gauge Non-Invariance
And Parity Nonconservation Of Three-Dimensional Fermions,'' Phys. Rev.
Lett. {\bf 52} (1984) 18.}
\nref\niemi{A. J. Niemi and G. W. Semenoff, ``Axial-Anomaly-Induced 
Fermion Fractionalization And Effective Gauge Theory Action In
Odd Dimensional Space-Times,'' Phys. Rev. Lett. {\bf 51} (1983) 2077.
}
\nref\ginsparg{L. Alvarez-Gaum\'e, S. Della Pietra, and G. Moore,
``Anomalies And Odd Dimensions,'' Ann. Phys. {\bf 163} (1985) 288.}
\nref\singer{I. M. Singer, ``Families Of Dirac Operators With
Applications To Physics,'' in {\it The Mathematical Heritage Of
Elie Cartan}, Asterisque (1985) 323.}
\refs{\redlich-\singer} for more detail)
a $\Z_2$ anomaly affecting the sign of
a Dirac determinant.  A case in point is the Rarita-Schwinger
operator $D$ of eleven-dimensional supergravity.  It has infinitely
many positive and negative eigenvalues, raising the question of
whether the sign of the fermion path integral can be consistently
defined.  In uncompactified eleven-dimensional Minkowski space $M^{11}$,
there is no problem; this follows from the absence of exotic
twelve-spheres, which implies that the diffeomorphism group of
$M^{11}$ is connected, so that there is no ``room'' for an anomaly.
It has not been clear under what circumstances the sign of the
Rarita-Schwinger determinant can be consistently defined after
compactification.

In field theory, when there is such a $\Z_2$ anomaly, it can generally
be canceled by including a Chern-Simons coupling $kL_{C.S.}$, where
$L_{C.S.}$ is a suitable Chern-Simons coupling that
is well-defined modulo $2\pi$, and $k$ is a {\it half}-integer
(congruent to $1/2$ modulo $\Z$), so that $e^{ikL_{C.S.}}$
 has a sign problem that just cancels
the anomaly of $\det D$.  The choice of a non-zero $k$, however,
violates parity, so this  effect is often described as a ``parity
anomaly.''  This has led to speculations that eleven-dimensional
supergravity or $M$-theory might have a parity anomaly, but this
would rule out its relation \hw\ to the $E_8\times E_8$ heterotic
string, which depends on dividing by an involution that reverses
the parity.  A parity anomaly of $M$-theory would also have strange
implications for other duality statements in string theory, as 
orientation-reversing 
transformations in $M$-theory are often mapped to known symmetries
of string theories.  

In section five, we resolve these questions.  Let $I_M
=CGG+{\rm gravitational~terms}$ be the  Chern-Simons
couplings in the low energy limit of $M$-theory.  
The $G$-dependence of $e^{iI_M}$ is
well-defined, according to our results in section four.  
However,  $e^{iI_M}$ has
a $G$-independent
sign ambiguity, which just equals that of the Rarita-Schwinger
determinant $\det D$, so that $e^{iI_M}\cdot\det D$ is well-defined.  
Since the other massless fields of $M$-theory are bosons with
manifestly well-defined determinants, this completes the
demonstration that the low energy effective action of $M$-theory
is well-defined.

Despite this crucial role of the Chern-Simons terms, there
is no parity violation.   The Chern-Simons coupling
that cancels the potential anomaly of the Rarita-Schwinger field is
completely parity-invariant, with  $C$ and $G$ understood (as usual)
to be odd under parity.  Roughly speaking, $G$ plays the role
that is usually played by the Chern-Simons
coupling $k$ in corresponding $8k+3$-dimensional
gauge theories.  When $\lambda$ is not divisible by two, one
must choose a  non-zero $G$, with half-integral $G/2\pi$
(just as in the field
theory one chooses a non-zero and half-integral $k$), and this violates
parity. But because $G$ is a field (while the usual
$k$ is a coupling constant) what one has in $M$-theory
is  spontaneous parity violation by the choice of a particular
physical state, rather than explicit parity violation of the theory.
Notice that, even though $G$ and $\lambda$ transform oppositely
under parity, the relation \unbu\ is completely parity-invariant;
in fact, as $\lambda$ is integral, it is equivalent to say that
$G/2\pi-\lambda/2$ is integral or that $G/2\pi+\lambda/2$ is integral.

\newsec{Flux Quantization}

\subsec{The Boundary Of The Universe}

The shifted quantization condition on $G$ that was described
in the introduction is easily motivated given the relation
\hw\ of $M$-theory to the $E_8\times E_8$ heterotic string.

Suppose that the space-time manifold $Y$ has a boundary $N$.
Then, according to \hw, there are $E_8$ gauge fields propagating
on $N$.  An $E_8$ bundle $V$ has a four-dimensional characteristic
class $w(V)$ which is associated with the differential form
 $\tr F\wedge F/16\pi^2$.  
The topology of $E_8$ is such that $w(V)$ is subject to absolutely
no restriction except for being integral.  The characteristic class
$\lambda=p_1/2$ of the tangent bundle of $Y$ is associated with the
differential form $\tr R\wedge R/16\pi^2$.  
In the second paper in \hw, it was determined
that the boundary values of $G$ are restricted by 
\eqn\uttu{\left.{G\over 2
\pi}\right|_N=
{1\over 16\pi^2}\left(\tr 
F\wedge F -{1\over 2}\tr R\wedge R\right).} 
At the level of cohomology, this 
implies that
\eqn\muttu{\left.\left[{G\over 2\pi}\right]\right|_N
=w(V)-{\lambda\over 2}}
and therefore, since $w(V)$ is integral, that
$[G/2\pi]-{\lambda/2}$ is integral when restricted to $N$.
So the promised relation \unbu\ holds, at least when restricted to $N$.

Now consider a physicist who is integrating $G/2\pi-\lambda/2$ over
a four-cycle $S$ in space-time.  If anywhere in the universe
-- maybe $10^{10} $ light years away -- there is a boundary $N$ that
contains a four-cycle $S'$ homologous to $S$, then from what
has just been said, the integral over $S$ of $G/2\pi-\lambda/2$ must
be integral.  If $\int_S(G/2\pi-\lambda/2)$ were to be non-integral,
the physicist measuring it would obtain the amazing information
that no $N$ and $S'$ exist, at any distance from $S$, no matter
how great.  The ability to obtain such information about what there
is at arbitrarily big distances from the observer 
seems counterintuitive, so it is natural to suspect that the right
requirement is
that $\int_S(G/2\pi-\lambda/2)$ is integral whether $N$ and $S'$ exist or
not, or in other words (since this is true for all $S$)
that the cohomology class $[G/2\pi]-\lambda/2$ is integral.
In the rest of this paper it will, hopefully, become clear that that
is the right interpretation.

For some additional insight, note that, since the homotopy
groups $\pi_i(E_8)$ vanish for $i<15$, except for $\pi_3$,
and since $\dim \, N<16$, there is no restriction on the
characteristic class $w(V)$ except integrality, and the $E_8$ bundle
$V$ over $N$ is completely determined (topologically) by $w(V)$.
The first statement means that \muttu\ imposes no restriction
on $G$ except integrality of the restriction to $N$ of
 $[G/2\pi]-\lambda/2$; this is the basis for an assertion
made above.  The second
statement means that, since $[G/2\pi]$ and $\lambda$ are determined
by their restrictions to the interior of space-time, a physicist who
has made thorough measurements of the physics away from the boundary
of space-time can uniquely predict what the $E_8$ bundle will have
to be on the boundary.

\subsec{Membrane Anomalies}

Now we will offer a second and I believe conclusive
approach to  the same result,
based on considering world-volume anomalies of $M$-theory membranes.
There is no need here to assume that $M$-theory has
``elementary'' membranes, whatever those may be.  It is only
necessary that $M$-theory admits macroscopic membranes.  Given this,
the path integral in the presence of the membrane must be well-defined;
this is the issue that we will examine.

Rather than aiming for generality, I will formulate the argument in
a representative situation.  Suppose the eleven-manifold $Y$ contains
a submanifold $D=\S^3\times \S^1$.  We consider a membrane whose
world-volume $T$ is wrapped over  $\S^3\times q$, where $q$ is
a point in $\S^1$.  We want to determine whether the membrane path
integral is single-valued when $q$ moves all the way around the $\S^1$.

There are two potentially dangerous factors to consider.  One is
the ``Chern-Simons'' factor coming from the coupling of the membrane
world-volume to $C$.  This factor gives in the membrane path integrand
a factor
\eqn\tugo{\exp( i\int_TC).}
The change in this factor when $q$ moves once around $\S^1$ is
\eqn\nugo{\exp(i\int_D G).}
This factor is 1 if and only if the period of $G$ is a multiple of $2\pi$.

The second dangerous factor is the path integral over the world-volume
fermions on $T$.  At this point, therefore, we need some remarks
on fermion path integrals.  The  path integral in $8k+3$ dimensions for 
massless fermions coupled to gauge fields and gravity
is naturally real, but can be positive or negative.  Potentially, as 
in \refs{\aw - \singer}, 
there can be an inconsistency in defining the sign
of the fermion path integral.  This will happen in our problem under some
conditions on the embedding of $D$ in $Y$.

In fact, the fermions on the membrane world-volume $T$ are sections
of the spin bundle of $T$ tensored with the normal bundle $N$ to $T$ in
space-time.  One has $N={\cal O }
\oplus N'$, where $\cal O$ -- a one-dimensional
trivial bundle -- is the direction tangent to the $\S^1$ factor in $D$,
and $N'$ is the normal bundle to $D$ in space-time.  $N'$ can be absolutely
any ${\rm Spin}(7)$ bundle.  Such a bundle has a four-dimensional
characteristic class $\lambda(N')$ (equal to $p_1(N')/2$). 
$\lambda(N')$ 
is equal to the restriction of $\lambda(Y)$ to $D$ (as the tangent bundle
to $D$ has vanishing characteristic classes).
The considerations of \refs{\aw - \singer} 
show that as $q$ moves once around
the $\S^1$, the sign of the fermion path integral changes by
a factor of $(-1)^{\int_D\lambda}$.  What must be single-valued is
therefore not really \nugo\ but
\eqn\onugo{(-1)^{\int_D\lambda}\exp(i\int_DG).}
But this amounts to our main claim: $G/2\pi$ has a half-integral period
on just those cycles on which the value of $\lambda $ is odd.

The fact that this computation gives the same result as the
argument based on boundaries of space-time and $E_8\times E_8$
gauge fields is a satisfying test of quantum $M$-theory.

\subsec{Application To An Orbifold}

\def\RP{{\bf RP}}
From what we have seen, it will sometimes happens that a
half-integral quantum of $G/2\pi$ will be trapped on a four-cycle
$D$ in space-time.  This is reminiscent of an example already
considered in the literature, where in $M$-theory 
on $\R^5/\Z_2$, a half-integral flux of $G$ appeared 
\ref\fivewitten{E. Witten, ``Five-Branes And $M$-Theory On An Orbifold,''
hepth/9512219.}.  To be precise, if $D$ is a four-cycle surrounding
the origin in $\R^5/\Z_2$ -- so $D$ can naturally be taken to be
a copy of $\S^4/\Z_2=\RP^4$ -- then 
it was found from considerations of space-time anomaly cancellation that
$\int_D(G/2\pi) = 1/2$ modulo $\Z$.

To put this in the present context, we should first delete a neighborhood
of the origin in $\R^5/\Z_2$ -- as field theory considerations
may not be applicable near the singularity.  Let $\tilde \R^5/\Z_2$
be $\R^5/\Z_2$ with such a neighborhood omitted.  Now, $\tilde\R^5/\Z_2$
is not orientable -- this is the only place in the present paper
where we consider an unorientable space-time.  The differential
form $\lambda$ can be taken
to vanish for $\tilde \R^5/\Z_2$ -- as that manifold
admits a flat metric.  So any statement we make will have to involve
torsion.

In any event, for unorientable manifolds, it is unnatural to state
a relation between $G$ and $\lambda$, as the two transform oppositely
under reversal of orientation (being respectively odd and even).
Note, though,\foot{The following formulation was improved by
suggestions by D. Freed.} 
that in the orientable case, $\lambda$ is congruent modulo
two to the Stieffel-Whitney class $w_4$, so that our statement in
the orientable case could equivalently have been
\eqn\doggo{\int_D {G\over 2\pi}={1\over 2}\int_D w_4\,\,\,
{\rm modulo}\,\,\Z.}
This is a way of expressing our result in the orientable case that
may carry over better to the unorientable case. 
Indeed, if as above $D$ is a copy of $\RP^4$ wrapping around the
``interior'' of $\tilde \R^5/\Z_2$, then a standard computation
\foot{The tangent bundle of $\tilde \R^5/\Z_2$ is a sum of
five copies of an unorientable real line bundle $\epsilon$. If
$x=w_1(\epsilon)$, 
then the total Stieffel-Whitney class of $\tilde \R^5/\Z_2$
is $(1+x)^5=1+x+x^4+x^5$.  In particular, $w_4(\tilde \R^5/\Z_2)=x^4$.
Now restrict this to a copy of $D=\RP^4\subset \tilde \R^5/\Z_2$.
The mod two cohomology of $\RP^4$ is the polynomial ring in $x$
with relation $x^5=0$, and in particular $x^4$ is the mod two
fundamental class of $\RP^4$, which integrates to one (modulo two).} 
shows that
\eqn\oboggo{\int_D w_4=1\,\,\,{\rm modulo}\,\,2.}
(Of course, as Stieffel-Whitney classes are mod two classes, the
integral only makes sense modulo two.)  Therefore, the  result of
\fivewitten\ that an $\R^5/\Z_2$ singularity is the source
of a half-integral flux of $G/2\pi$ is actually a consequence of
\doggo.  If one considers other somewhat analogous $\Z_2$ orbifold
singularities studied in $M$-theory, such as $\R^8/\Z_2$ and
$\R^9/\Z_2$ considered in \ref\mukhi{K. Dasgupta and S. Mukhi,
``Orbifolds Of $M$-Theory,'' hepth/9512196.}, a computation
as in the footnote shows that $w_4=0$, so that no half-integral $G$ flux
is expected - or found.

\newsec{Integrality Of The Number Of Branes}

In \svw, compactification of $M$-theory on an eight-manifold
$X$ was considered.  It was tacitly assumed that the vacuum expectation
value of $G$ was  zero.  The effects of the interaction
\refs{\vw,\duff}  $C\wedge I_8(R)$ (where $I_8(R)$ is
a certain quartic polynomial in the curvature tensor) were considered.
If $J=-\int_XI_8(R)$ is non-zero, one obtains a ``tadpole'' for the $C$ field,
which must be canceled by including in the vacuum $J$ two-branes.
Since the number of two-branes must be an integer, the theory
is inconsistent unless $J$ is integral.

It was pointed out in \svw\ that for $X$ a Calabi-Yau four-fold,
$J=c_4/24$ (or equivalently $J=\chi(X)/24$, with $\chi$ the topological
Euler characteristic).  Index theorems were used to show that
$c_4$ is always divisible by six, but explicit examples show that
$c_4$ need not be divisible by 12 or 24, so that one gets
a non-trivial restriction on these compactifications.

For a more general eight-manifold, the formula for $J$, if written
in terms of $\lambda=p_1/2$ and $p_2$, is
\eqn\formj{J={p_2-\lambda^2\over 48}.}
Now let us assess the integrality of $J$, using index theorems
together with our previous considerations.
The index of the Dirac operator on $X$, written in terms of $\lambda$
and $p_2$, is 
\eqn\normj{I={1\over 1440}\left(7\lambda^2-p_2\right).}
Since $I$ is an integer, it follows that
\eqn\ormj{p_2-\lambda^2\cong 6\lambda^2 \,\,{\rm modulo}\,1440.}

Now, if $G$ is to vanish, then according to our shifted flux condition,
$\lambda$ must be divisible by two, say $\lambda=2x$ with
$x$ an integral class.  So we can write
\eqn\normj{p_2-\lambda^2\cong 24 x^2 \,\,{\rm modulo}\,1440.}
So $p_2-\lambda^2$ is divisible by 24.
Since we need divisibility by 48, we must probe more deeply.

An extra factor of two arises as follows.  
Let $x$ be any element of $H^4(X,\Z)$.  Then by a special case of
the Wu formula,
\eqn\offo{x^2\cong x\cdot \lambda \,\,{\rm modulo}\,\, 2.}  
If, therefore, $\lambda$ is divisible by two, then $x^2$ is even,
and \normj\ implies that $(p_2-\lambda^2)/48$ is integral, showing
that the number of branes is integral, as promised.
Note that formula \offo\ implies
that the intersection form on $H^4(X,\Z)$ is even when $\lambda$ is
divisible by two.  Since this intersection form is in any case unimodular
(by Poincar\'e duality), this gives another occurrence of even unimodular
lattices in string theory.
     
The formula \offo\ can be put in the following theoretical context.
Let $x$ and $y$ be elements of $H^4(X,\Z)$.  Because $(x+y)^2$  
is congruent to $x^2+y^2$ modulo two, the function $f(x)=x^2$
is a linear function of $x$ modulo two. By Poincar\'e duality,
there is therefore an element $\alpha\in H^4(X,\Z_2)$ such that for all
$x\in H^4(X,\Z_2)$, $x^2\cong x\cdot \alpha$ modulo 2.  
$\alpha$ is called
the Wu class and in general can be written as a polynomial
in Stiefel-Whitney classes.  For $X$ a spin manifold, so
that $w_i=0$ for $i<4$, one has simply $\alpha=w_4$, and this in turn 
equals the mod two reduction of the class $\lambda=p_1/2\in H^4(X,\Z)$,
leading to \offo.  We will not attempt here an account
of these matters, but instead give a direct proof of \offo, using
$E_8$ index theory, in the next section.

In a similar way, because the function $f(x)=x^3$ is linear in $x$
modulo 3 (that is, $(x+y)^3\cong x^3+y^3$ modulo three), it is natural, if
$x$ is for example a four-dimensional class in a twelve-dimensional
spin manifold $W$, to have a formula $x^3\cong 
x\cdot \beta$ modulo 3, with
$\beta$ some eight-dimensional class.  In the next section, we will
explain how the existence of such a formula is relevant to $M$-theory,
and we will determine $\beta $ using $E_8$ index theory.

The computation we did in this section was really a special
case of the general question of whether the Chern-Simons
couplings -- the classical $CGG$ coupling and the quantum correction
$CI_8(R)$ -- are single-valued.  Indeed, non-integrality of the tadpole
would mean that the $CI_8(R)$ coupling is not well-defined modulo
$2\pi$ upon adding to $C$ a closed form with properly normalized
periods.  In the rest of this paper, we study more systematically
the well-definedness of the Chern-Simons couplings.

\newsec{The Chern-Simons Couplings}

Since $G$ can have non-zero periods, the definition $G=dC$ is not
really valid globally, and $C$ is not well-defined as a differential
form.  There is therefore some subtlety in defining the classical
Chern-Simons interaction $I=\int_YC\wedge G\wedge G$.  This is most
naturally accomplished by realizing $Y$ as the boundary of a
twelve-dimensional spin manifold
$Z$, extending the closed four-form $G$ over $Z$,
\foot{The existence of a twelve-dimensional spin
manifold $Z$ over which $G$ extends
is ensured by a computation by Stong \ref\stong{R. Stong,
``Calculation Of $\Omega^{\rm spin}_{11}(K(\bf Z,4))$,'' in
{\it Unified String Theories}, eds. M. B. Green and D. J. Gross
(World Scientific, 1986).}.   For later purposes, we want to choose
the extension of $G$ to obey the shifted quantization condition.  $Z$ can
be chosen to make this possible.  
  For instance, the shifted quantization condition
says that $[G/\pi]=\lambda+2\alpha$ where $\alpha$ is some integral
class.  $\lambda$, being a characteristic class,
automatically extends over $Z$, and Stong's
theorem means that $Z$ can be chosen so that $\alpha$ extends;
one can then take $G/\pi=\lambda+2\alpha$ as the definition of the
extension of $G$.}  
and setting $I=\int_Z G\wedge G\wedge G$.  To check whether this
is well-defined -- that is, independent of the choice of $Z$ and of
the extension of $G$ -- one takes two different choices, involving
twelve-manifolds $Z$ and $Z'$, each with boundary $Y$, and one tries
to show that the corresponding $I$'s -- call them $I_Z$
and $I_{Z'}$ -- are equal modulo $2\pi$.  If $Q$ is the closed
twelve manifold obtained by gluing $Z'$ to $Z$ along their common
boundary (one takes $Z'$ and $Z$ with ``opposite'' orientations
so that the orientations fit together to an orientation of $Q$),
then 
\eqn\uttor{I_Z-I_{Z'}=\int_{Q}G\wedge G\wedge G.}
For the Chern-Simons interaction to be well-defined modulo
$2\pi$ is thus equivalent to the requirement
that for a closed spin manifold $Q$, $I_Q=\int_QG\wedge G\wedge G$
is a multiple of $2\pi$.  

One might expect that the quantization law of $G$ would be
such as to ensure this, but in fact it is not.
The actual relation is that, if
$\alpha$ is the cohomology class of $[G/2\pi]$, then
\eqn\hillo{{I_Q\over 2\pi}=-{1\over 6}\int_Q\alpha^3.}
We will see below where this crucial factor of $-1/6$ comes from
in terms of the relation of $M$-theory to the heterotic string.
It can also be verified using a recent careful analysis of
numerical factors in $M$-theory \ref\dealwis{S. De Alwis,
``Anomaly Cancellation In $M$-Theory,'' hepth/9609211.}.

Since $1/6=1/2-1/3$, in analyzing integrality of the right hand
side of \hillo,
mod 2 and mod 3 congruences are helpful.  The mod 2 congruences
one needs are generalizations of \offo, and the mod 3 congruences
are possible for a reason suggested toward the end of section three.
In addition to certain congruences, restoring integrality depends 
on gravitational corrections to the $CGG$ coupling, which are of
two kinds. (1) It is not $\alpha$, but $\alpha-\lambda/2$, that is
integral.  (2) One must also include the $CI_8(R)$ quantum correction.

Including the gravitational corrections and working out all of the
congruences that are relevant in understanding the integrality of
$I_Q/2\pi$ would be extremely complicated in the absence of some insight
about the structure that is appearing here.    But the
analysis is actually made extremely  easy by the recognition in
\hw\ that the Chern-Simons couplings in eleven dimensions are related
to anomalies in $E_8$ supergravity in ten dimensions.  They are therefore
also related to index theory in twelve dimensions.  To  be more
exact, the eleven-dimensional Chern-Simons couplings were related
in \hw\ to the anomalies of gluinos
of {\it one} $E_8$ gauge group (rather
than $E_8\times E_8$) plus {\it half} of the anomaly of the dilatino
and gravitino.  Moreover, the gluinos, gravitino, and dilatino are
all Majorana-Weyl, so their anomalies are half of what one
would have for ``ordinary''  Weyl fermions.

\nref\oldalwis{S. de Alwis,
``A Note On Brane Tension And $M$-Theory,'' hepth/9607011.}
Introduce on the twelve-manifold $Q$ an $E_8$ bundle $V$ whose
characteristic class $w(V)$ obeys $w=\alpha+\lambda/2$.  ($V$ exists
and is unique as explained at the end of section 2.1.)  The shifted
flux condition permits $w$ to be an arbitrary four-dimensional class.
Let $i(E_8)$ be the index of the Dirac operator on $Q$ for fermions
with values in $V$ (taken in the adjoint representation of $E_8$),
and let $i(R.S.)$ be the index of the Rarita-Schwinger operator on $ Q$.
The analysis of the Chern-Simons terms in section 3.1 of \hw,
as subsequently extended \refs{\dealwis,\oldalwis},
is equivalent
to the assertion that after including the gravitational corrections
\eqn\jello{{I_Q\over 2\pi}={i(E_8)\over 2}+{i(R.S.)\over 4}.}
Here $i(E_8)$ appears with a factor of $1/2$ because of the Majorana-Weyl
condition, while $i(R.S.)$ has a factor of $1/4$ because of the 
Majorana-Weyl condition plus the fact that we want the characteristic
class that is related to $1/2$ of the gravitino-dilatino anomaly.

Now, in $8k+4$ dimensions, because of charge conjugation symmetry,
the index of the Dirac operator with values in a real vector
bundle is even.  So $i(E_8)/2$ is an integer, but $i(R.S.)/4$ is in
general a half-integer.   Since the $G$ or $w$ dependence of
$I_Q/2\pi$ is entirely in $i(E_8)$ (the other term is not sensitive
to the choice of $E_8$ bundle!), $I_Q/2\pi$ changes by an integer
under any change of $G$ or $w$.  Thus, we have verified integrality of
$I_Q/ 2\pi$ up to a term that is $G$-independent.  Moreover,
in general \jello\ makes it clear that $I_Q/2\pi$ takes values
in $\Z/2$, with integrality depending only on the value of $i(R.S.)$
modulo four.  In the next section, we will interpret the meaning
of this last, $G$-independent, potential failure of integrality.

To exhibit the congruences that are implicit here,
and also to make clear just how much trouble one might have had with
this analysis if one did not know the relation of $M$-theory to 
$E_8$ gauge theory, 
we will now expand \jello\ explicitly using the index theorem
(or equivalently, the detailed knowledge of the various Chern-Simons
terms) and make explicit {\it some} of the relevant congruences.  We get
\eqn\omello{{I_Q\over 2\pi}=-{1\over 6} \int_Q\left(w-\half \lambda\right)
\left((w-\half \lambda)^2-{1\over 8}(p_2-\lambda^2)\right).  }
(Notice that $I_Q/2\pi=-(1/6)\int_Qw^3$ modulo gravitational
corrections; as $w$ equals
$\alpha$ modulo a gravitational correction, this is the basis
for the assertion that $I_Q/2\pi$ is as given in \hillo\ modulo
gravitational corrections.) 
Let us  see what  congruences can be extracted from the
knowledge that the $w$-dependent terms are integral.
First we consider mod 3 congruences.  

After expanding out
the right hand side of \omello\ and dropping terms that do not
have a 3 in the denominator, the fact that the $w$-dependent part
of $I_Q/2\pi$  is a 
3-integer (it can be written as a rational number with denominator
not divisible by three) turns out to be equivalent to the assertion that
\eqn\jumello{w^3\cong -(p_2-\lambda^2)w ~~{\rm modulo}~3.} 
Such a congruence was promised at the end of section three, and
enables one to reexpress $w^3/6$ (which appears 
in \hillo) in terms of $w^3/2$
and a gravitational correction, linear in $w$.   If \jumello\ were known
independently, this would be a step in the direction of analyzing
integrality of \hillo\ without using $E_8$   gauge theory.

For an interesting mod 2 congruence, take $Q=\S^1\times Y$, with
$Y$ an eleven-manifold, and take $w=d\theta\cdot u+v$, where
$d\theta$ is a closed one-form on $\S^1$ that integrates to 1,
$u\in H^3(Y,\Z)$, and $v\in H^4(Y,\Z)$.  Think of $I_Q/2\pi$
as a function of $u$ and $v$.  Since this function changes
by an integer when $w$ is changed, $T(u,v)=(I_Q(u,v)-I_Q(u,0)-I_Q(0,v)
+I_Q(0,0))/2\pi$ is an integer.  But concretely
\eqn\throgh{T(u,v)=-{1\over 2}\int_Y\left(uv^2-uv\lambda\right).}
Integrality of this expression for arbitrary $u$ implies that
\eqn\offix{v^2\cong v\lambda ~{\rm modulo}~ 2}
for any four-dimensional class $v$ on an eleven-manifold $Y$.  
If we specialize to the case that $Y=\S^3\times X$, with $X$ an
eight-dimensional spin manifold and $v\in H^4(X,\Z)$, then
\offix\ reduces  to \offo, giving the promised proof of \offo\
based on $E_8$ index theory.

For one final comment, suppose that $\lambda(Q)$ is even.
Then we can choose an $E_8$ bundle on $Q$ with  $w=\lambda(Q)/2$. For
such a bundle, $I_Q$ is clearly zero according to \omello.  Hence,
using \jello, $i_{R.S.}$ is divisible by four when $\lambda$ is even.
Since that is not so in general, we go on in the next section to
analyze the significance of the fact that $I_Q/2\pi$ may be half-integral.

Note that the formula \omello\ is completely invariant under reversal
of orientation of $Q$ (which changes the sign of the index) together
with $w\to\lambda-w$.  Of course $w\to\lambda-w$ corresponds to $G\to -G$,
which customarily accompanies orientation reversal in $M$-theory.

\newsec{ The Rarita-Schwinger Path Integral }

We still must interpret the fact that $I_Q/2\pi$ is half-integral
in general.  On the other hand, there is one more potentially anomalous
factor in the low energy effective action of $M$-theory.  This is the
determinant of the Rarita-Schwinger operator $D$.  We will see that the
two problems cancel each other.

As we have already noted in section 2.2, in $8k+3$ dimensions  a 
massless 
fermion path integral is naturally real.  This is 
essentially because the  massless
Dirac operator is hermitean and has real  eigenvalues.  But (for Majorana
fermions) the path integral has no natural
sign, because there is no natural way to fix the sign of the fermion
measure.  

One way, as in \aw, to try to define the determinant of the 
Rarita-Schwinger operator $D$ is to compare $\det D$ to the determinant 
of a massive Rarita-Schwinger operator $D_m=D+im$.  The constant $m$ is 
odd under parity, as a result of which $\det D_m$ is complex.
Since the mass is a soft perturbation and one can define $\det D$ and 
$\det D_m$ using the same fermion measure, the ratio $\det D/\det D_m$
is completely well-defined.  Then by taking the limit as $m\to\infty$,
one comes, in a sense,
as close as one can to getting a natural definition of $\det D$
that preserves all the formal properties.  However, one 
gets different limits by taking $m\to +\infty$ or $m\to -\infty$,
so that the parity violation which follows from a choice of sign of $m$ 
survives as $|m|\to\infty$.  The limits are in fact essentially (up to a
sign that depends only on the manifold and not on the metric)\foot{
A more precise result in which also this overall manifold-dependent sign
is fixed 
can be obtained by using the eta invariant instead of Chern-Simons,
as in \refs{\ginsparg,\singer}.  This refinement is not important
in studying $M$-theory on any given manifold, but would be important in
comparing $M$-theory on different manifolds, for instance in
order to analyze gravitational instantons.}
\eqn\polimpo{\det D \exp(\pm iI_{R.S.}/2).}
Here the sign of the exponent depends on the sign of $m$.
Also $I_{R.S.}$ is a properly normalized Chern-Simons functional, 
associated
with the characteristic class $i(R.S.)/2$.  (We recall that $i(R.S.)$, the
Rarita-Schwinger index in $8k+4$ dimensions, is even.)  Because of
the $1/2$ in the exponent, the exponential factor in \polimpo\ may
change sign under some orientation-preserving diffeomorphism, but any
such sign changes precisely cancel sign changes in $\det D$, as one can
verify using index theory.  

This has led some physicists to suspect over the years that there
might be a quantum correction to 
the low energy effective action of $M$-theory
of the form $iI_{R.S.}/2$, or perhaps $-iI_{R.S.}/2$, or some other
expression that differs from these by properly normalized Chern-Simons
terms, to cancel the potential sign anomaly of $\det\,D$.
\foot{For example, see a brief comment in \ref\otherduff{M. J. Duff,
``$E_8\times SO(16)$ Symmetry of $d=11$ Supergravity?'' in
{\it Quantum Field Theory And Quantum Statistics}, eds.
I. A. Batalin, C. J. Isham, and G. A. Vilkoviski
(Adam Hilger, Ltd., Bristol, 1987), vol. 2., p, 209.}.}
  At first sight, it seems that any such functional would violate
the parity-invariance of $M$-theory, even if one allows the freedom to
add terms that depend on $G$ as well as the metric.  
Parity allows only terms odd in $C$ and $G$ (namely our friends
$CGG$ and $CI_8(R)$); how can these terms, which vanish at $G=0$, help
with a problem -- the sign of $\det D$ -- that persists at $G=0$?
But the shift in
the quantization law of $G$ that we have uncovered makes it just possible
to find a parity-invariant Chern-Simons-like interaction whose difference
from $I_{R.S.}/2$ is properly normalized.  The interaction with this
property is our friend, the Chern-Simons interaction of $M$ theory,
schematically $I_M=CGG+CI_8(R)$.  This is clear from \jello: 
$I_Q/2\pi$ differs from $i(R.S.)/4$ by an integer, so $I_M$, 
which is the Chern-Simons coupling derived from $I_Q/2\pi$, differs
from $I_{R.S.}/2$ by a properly normalized Chern-Simons coupling
(derived from $i(E_8)/2$).  The ability to ``have our cake and eat it
too,'' to cancel the sign ambiguity of $\det D$ with a Chern-Simons
interaction and still maintain parity conservation, depends on the fact
that, as $G$ must be congruent to $\lambda/2$ mod two, one cannot
set $G$ to zero (unless $\lambda$ is even, in which case, as we saw
at the end of section four, $i(R.S.)$ is divisible by four and the
problem vanishes).
Related to this, symmetry under reversal of orientation is rather
hidden in \jello; it acts in the peculiar fashion $w\to \lambda-w$
(which is manifest when \jello\ is expanded to get \omello).
Of course that transformation law comes from the classical parity
transformation law $G\to -G$ of $M$-theory, together with the shift
in the flux quantization law of $G$.

\bigskip
\bigskip
I am grateful to D. Freed and J. Morgan for helpful explanations
of some topological points.

\listrefs

\end